\documentclass[12pt]{article}
\usepackage{amsfonts,amsbsy}
\usepackage[dvips]{graphicx,color}
\usepackage{psfrag}
\usepackage{bm}
\usepackage[titletoc, title]{appendix}
\newtheorem{proposition}{Proposition}[section]

\newcommand{\up}{\uparrow}
\newcommand{\dn}{\downarrow}

\newcommand{\rme}{\mathrm{e}}
\newcommand{\im}{\mathrm{i}}
\newcommand{\Ne}{{N_\mathrm{e}}}
\newcommand{\Np}{{N_\mathrm{p}}}

\newcommand{\PhiG}{\Phi_\mathrm{G}}
\newcommand{\PhiGe}{\Phi_\mathrm{G,0}}
\newcommand{\PhiGo}{\Phi_\mathrm{G,1}}

\newcommand{\La}{\Lambda}
\newcommand{\barLa}{\bar{\Lambda}}
\newcommand{\cs}[1]{c_{#1,\sigma}}
\newcommand{\csd}[1]{c_{#1,\sigma}^\dagger}
\newcommand{\ns}[1]{n_{#1,\sigma}}
\newcommand{\nup}[1]{n_{#1,\uparrow}}
\newcommand{\ndn}[1]{n_{#1,\downarrow}}
\newcommand{\cd}[1]{c_{#1}^\dagger}

\newcommand{\as}[1]{a_{#1,\sigma}}
\newcommand{\asd}[1]{a_{#1,\sigma}^\dagger}
\newcommand{\aup}[1]{a_{#1,\uparrow}}
\newcommand{\adn}[1]{a_{#1,\downarrow}}
\newcommand{\aupd}[1]{a_{#1,\uparrow}^\dagger}
\newcommand{\adnd}[1]{a_{#1,\downarrow}^\dagger}
\newcommand{\tas}[1]{\tilde{a}_{#1,\sigma}}
\newcommand{\tasd}[1]{\tilde{a}_{#1,\sigma}^\dagger}
\newcommand{\tadL}{\tilde{a}_L^\dagger}

\newcommand{\utadL}{\underline{\tilde{a}}_L^\dagger}

\newcommand{\tad}[1]{\tilde{a}_{#1}^\dagger}
\newcommand{\taupd}[1]{\tilde{a}_{#1,\up}^\dagger}

\newcommand{\bs}[1]{b_{#1,\sigma}}

\newcommand{\bup}[1]{b_{#1,\uparrow}}
\newcommand{\bdn}[1]{b_{#1,\downarrow}}
\newcommand{\bupd}[1]{b_{#1,\uparrow}^\dagger}
\newcommand{\bdnd}[1]{b_{#1,\downarrow}^\dagger}
\newcommand{\bd}[1]{b_{#1}^\dagger}

\newcommand{\expece}[1]{\langle #1 \rangle_0}
\newcommand{\expeco}[1]{\langle #1 \rangle_1}
\newcommand{\expecl}[1]{\langle #1 \rangle_l}
\newcommand{\eqref}[1]{(\ref{#1})}
\begin{document}
\baselineskip=1.5\baselineskip
\begin{center}
\textbf{\Large 
 One-dimensional extended Hubbard model with 
 spin-triplet pairing ground states 
}\bigskip\\
Akinori TANAKA\footnote{akinori@ariake-nct.ac.jp}\bigskip\\
\textit{Department of General Education, National Institute of
 Technology,
Ariake College,
 Omuta, Fukuoka 836-8585, Japan}
\end{center}
\vspace*{4cm}
\begin{abstract}
\baselineskip=1.5\baselineskip
We show that the one-dimensional extended Hubbard model
has saturated ferromagnetic ground states with
the spin-triplet electron pair condensation 
in a certain range of parameters.
The ground state wave functions with fixed electron numbers 
are explicitly obtained.
We also construct two ground states in which both the spin-rotation 
and the gauge symmetries are broken, 
and show that these states are transferred from one to the other
by applying the edge operators.
The edge operators are reduced to the Majorana fermions in a special case. 
These symmetry breaking ground states 
are shown to be 
stabilized by a superconducting mean field Hamiltonian 
which is related to the Kitaev chain 
with the charge-charge interaction. 
\end{abstract}
\newpage
\section{Introduction}
The extended Hubbard model
has been studied extensively
to understand phenomena such as
charge density wave, spin density wave and unconventional
superconductivity
which can not be described by the Hubbard model
consisting of 
the electron hopping term 
and the on-site interaction term~\cite{Hirsch93,JK99,TML07}.
The Hamiltonian of the model is obtained by adding
interaction terms 
of electrons on different sites 
to the Hubbard Hamiltonian.
In the case where the added interaction together with the on-site one 
is dominant and is known to induce a certain ordering state with an energy gap, 
the model is well understood
by considering the electron hopping as a perturbation.
On the other hand, in order to understand phenomena 
which do not arise directly from
interactions, we have to face the difficult problem of analyzing the interplay 
between the electron hopping and some interactions in a convincing way.
The unconventional superconductivity corresponds to such a case. 

Here we restrict ourselves to the one-dimensional 
extended Hubbard model with
nearest neighbour interactions.
Despite the difficulty in analyzing correlated electron systems, 
there are a few rigorous results associated with superconductivity
in this case.
Most of these results are obtained through the Bethe ansatz method, and 
the superconducting ground states so far obtained 
are related to spin-singlet electron pair condensation~\cite{EKS92,AA94,DM01}.
In this paper we provide another rigorous result for the model.
By using a similar method in Ref.~\cite{Tanaka08},
we will show that the model exhibits 
saturated ferromagnetic, spin-triplet electron pair condensation 
in the ground state over a certain range of interaction parameters.

It is worth noting that in the last decade
the Majorana edge state formed on a spinless superconducting wire
has attracted much interest
both theoretically and 
experimentally~\cite{Kitaev01,Fendley12,LSS10,Alicea11,Mourik12,Lee14,Nadj14}.
Our model exhibits saturated ferromagnetism where the electrons behave 
as spinless fermions.
We show that a similar edge state is formed
in the gauge symmetry breaking ground state of our model.

This paper is organized as follows.
In Section~\ref{s:definition}, we give the definition and state the main result.
Section~\ref{s:Proof} is devoted to the proof of the main result.
In Section~\ref{s:GSwithBrokenSymmetries}, we introduce 
two ground states with broken spin-rotation and gauge symmetries, 
and show that these ground states exhibit similar properties to 
those of the spinless superconducting wire in the topological phase.
In Section~\ref{s:MeanField}, we consider mean fields which stabilize
the ground states introduced in the previous section.
In Section~\ref{s:ElectronNoConserving}, we investigate the properties 
of the ground state with the fixed number of electrons.
In Section~\ref{s:Anisotropy}, we extend the model to the case of
the anisotropic spin-spin interaction.
Finally, in Section~\ref{s:Conclusion}, we provide conclusions. 

\section{Definition of the model and the main result}
\label{s:definition}
We consider a one-dimensional array of $L$ sites, which are labeled as
$1,2,\dots,L$. 
We write $\La$ for the set of numbers $1,2,\dots,L$ and identify $\La$ 
with the array of $L$ sites. 
We also write $\barLa$ for $\La\backslash\{L\}$.
In this paper $L$ is assumed to be an odd integer with $L\ge3$.
This condition is adopted only for simplicity, and 
similar results for even $L$ are obtained with minor changes.

Let $c_{x,\sigma}(c_{x,\sigma}^\dagger)$ be the annihilation(creation)
operator of an electron at site $x\in\La$ and with spin $\sigma=\up,\dn$.   
They satisfy the anticommutation relations,
\begin{equation}
\{c_{x,\sigma},c_{y,\tau}\}
=
\{\cd{x,\sigma},\cd{y,\tau}\}
=0
\end{equation}
and
\begin{equation}
\{\cd{x,\sigma},c_{y,\tau}\}=\delta_{x,y}\delta_{\sigma,\tau}
\end{equation}
for any sites $x,y$ and any $\sigma,\tau=\up,\dn$.
For each site $x$, we define the  number operators $\ns{x}=\csd{x}\cs{x}$ and
$n_{x}=n_{x,\up}+n_{x,\dn}$, and the spin operators
$\displaystyle 
 S_{x}^{(l)}=\frac{1}{2}\sum_{\sigma,\tau}c_{x,\sigma}^\dagger p_{\sigma,\tau}^{(l)}
 c_{x,\tau}
$  
with $l=1,2,3$, where $p^{(l)}_{\sigma,\tau}$ are the elements of
the Pauli matrices
\begin{equation}
 p^{(1)}
  =\left(
    \begin{array}{@{\,}cc@{\,}}
     0 & 1 \\
     1 & 0 
    \end{array}
	  \right),~
 p^{(2)}
  =\left(
    \begin{array}{@{\,}cc@{\,}}
     0 & -\im \\
     \im & 0 
    \end{array}
	  \right),~
 p^{(3)}
  =\left(
    \begin{array}{@{\,}cc@{\,}}
     1 & 0 \\
     0 & -1 
    \end{array}
	  \right).
\end{equation}

For each nearest neighbour pair of sites $x$ and $x+1$, we define 
local Hamiltonian $H_{x}$ by
\begin{equation}
 H_x =
       H_{t,x}+H_{U,x} +H_{V,x}+H_{J,x}+H_{X,x}
\end{equation}
where
\begin{eqnarray}
 H_{t,x} 
   & = & -t\sum_{\sigma=\up,\dn}
   (\csd{x}\cs{x+1}+\csd{x+1}\cs{x})
             - \mu_x n_{x} - \mu_{x+1} n_{x+1},\\
 H_{U,x}   & = & U (\nup{x}\ndn{x}+\nup{x+1}\ndn{x+1}),
\label{eq:on-site}
\\
 H_{V,x}   & = & -V n_x n_{x+1}, \\
 H_{J,x}   & = & J\left(\frac{n_x n_{x+1}}{4}-
                              \bm{S}_x\cdot \bm{S}_{x+1}\right),
\label{eq:spin-spin}
			\\
 H_{X,x} & = & \sum_{\sigma=\up,\dn}
                     (X_{x}n_{x,-\sigma}+X_{x+1}n_{x+1,-\sigma})
		     (\csd{x}\cs{x+1}+\csd{x+1}\cs{x}).
\end{eqnarray}
The term  $H_{t,x}$ represents electron hopping, and 
$H_{U,x}, H_{V,x}, H_{J,x}$ and $H_{X,x}$ 
represent  electron-electron interactions, usually referred to as
the on-site, the charge-charge, the spin-spin
and the bond-charge interactions, respectively.  
In this paper, we assume $0<2t\le V$ and 
define parameter $\delta$ ranging from $0$ to $\pi/2$ by
\begin{equation}
 \sin \delta =\frac{2t}{V}.
\end{equation} 
We then consider the Hamiltonian given by
\begin{equation}
 H=\sum_{x\in\barLa} H_x
\end{equation}
on $\La$ with open boundary conditions.

Before stating our main result, we have to introduce some more
notations.
Let us define $\tilde{a}$ operators by
\begin{equation}
 \tas{x}=
  \left\{
   \begin{array}{@{\,}ll}
    \displaystyle \frac{1}{\sin\delta}\left(\sum_{y=1}^x w_y \cs{y}
                   - \sum_{y=x+1}^L w_y \cs{y}\right)
    & \mbox{if $x\in\barLa$}; \bigskip\\
    \displaystyle \frac{1}{\sin\delta}\sum_{y=1}^L w_y\cs{y} 
     &\mbox{if $x=L$}
   \end{array}
  \right.
\label{eq:ta}
\end{equation}
where 
\begin{equation}
w_x=
\left\{
 \begin{array}{@{\,}ll}
  \sin(\delta/2) & \mbox{if $x$ is odd}; \\
  \cos(\delta/2) & \mbox{otherwise.}
\end{array}
\right.
\end{equation}
By using the $\tilde{a}$ operators, we define pair operators
$\zeta_{\sigma,\tau}^\dagger$ with $\sigma,\tau=\up,\dn$ by
\begin{equation}
 \zeta^\dagger_{\sigma,\tau}
 =
 \sum_{x,y\in\La} {\sf F}_{x,y}
             \tilde{a}^\dagger_{x,\sigma}
	     \tilde{a}^\dagger_{y,\tau}, 
\label{eq:zeta}
\end{equation}
where ${\sf F}_{x,y}$ is given by
\begin{equation}
 {\sf F}_{x,y}=
  \left\{
   \begin{array}{@{\,}ll}
    -\frac{1}{2}\sin\delta & \mbox{if $y-x=1$};\bigskip\\
    -\frac{1}{2}\sin\delta & \mbox{if $x=1,y=L$};\bigskip\\
    \frac{1}{2}\sin\delta  & \mbox{if $x-y=1$};\bigskip\\
    \frac{1}{2}\sin\delta  & \mbox{if $x=L,y=1$};\bigskip\\
    0  & \mbox{otherwise}.
   \end{array}
  \right.
\label{eq:Fxy}
\end{equation}
It is noted that ${\sf F}_{x,y}=-{\sf F}_{y,x}$.

We denote by $\Phi_0$
the state with no electrons on $\La$.
The total number of electrons on $\La$ is denoted by $\Ne$.
We assume $0\le \Ne \le L$ and define the number $\Np$
of electron pairs by
\begin{equation}
\Np=
  \left\{
   \begin{array}{@{\,}ll}
    \frac{\Ne}{2} & \mbox{for even $\Ne$};\bigskip\\
    \frac{\Ne-1}{2} & \mbox{for odd $\Ne$}.\\
   \end{array}
 \right.
\end{equation}

With the values of the parameters given by 
\begin{eqnarray}
U_0 & = &  V\sin^2\delta = \frac{4t^2}{V}, \\
J_0 & = &  V(2-\sin^2\delta)=2V-\frac{4t^2}{V}, \\
X_0 & = &  \frac{V}{2}\sin\delta\cos\delta
=t\sqrt{1-\left(\frac{2t}{V}\right)^2},
\end{eqnarray}
our main result is summarized as follows:
\begin{proposition}
\label{proposition}
 Suppose that both $U>U_0$ and $J\ge J_0$ are satisfied. 
Then, the ground state energy of $H$ with 
\begin{equation}
 X_x =  (-1)^{x+1}X_0 \label{eq:condition_X}
\end{equation}
and
\begin{equation}
  \mu_x=
  -\frac{V}{2}\left\{1-(-1)^x\cos\delta\right\}
  \label{eq:condition_mu}
\end{equation}
is zero for $0\le \Ne \le L$.
For fixed $\Ne$, the ground state is unique apart from the degeneracy
due to the spin-rotation symmetry, 
and is given by
\begin{equation}
\PhiG=
  \left\{
   \begin{array}{@{\,}ll}
     \left(\zeta^\dagger_{\up,\up}\right)^\Np\Phi_0 
      & \mbox{for even $\Ne$} \\
     \taupd{L}\left(\zeta^\dagger_{\up,\up}\right)^\Np\Phi_0 
      & \mbox{for odd $\Ne$} \\
   \end{array}
   \right.
\label{eq:ground_states}
\end{equation}
and its $SU(2)$ rotations.
\end{proposition} 

The parameters in the Hamiltonian $H$ must satisfy several conditions 
so that $\PhiG$ can become the ground state of $H$.
Here we briefly comment on the physical feasibility of these conditions.
Firstly we need sufficiently large on-site repulsion and nearest neighbour 
ferromagnetic interaction.
These are necessary mainly to stabilize the ferromagnetic state.
(Note that the Hamiltonian $H$ is proved to exhibit metallic ferromagnetism
for $J>0$ in the limit $U\to\infty$~\cite{Tasaki98}.)
Ferromagnetic materials are expected to satisfy these conditions.
As for the nearest neighbour charge-charge interaction, it must be attractive.
Although the charge-charge interaction arising directly 
from the Coulomb interaction is repulsive,
it may become an effective attractive interaction 
such as a phonon-mediated interaction.
The strength $V$ of the charge-charge interaction also needs to be $2t\le V$.
This condition will hold in systems with narrow conduction bands.
We furthermore need to fine-tune $\mu_x$ and $X_x$.
Note, however, that the anisotropic spin-spin interaction removes the condition
on $X_x$ (see Proposition~\ref{proposition2}).
Although it will be difficult to fine-tune $\mu_x$ and $X_x$,  
the ground state $\PhiG$ or a state which has a large overlap with $\PhiG$ may be realized 
in one-dimensional ferromagnetic materials with narrow conduction bands 
if an effective attraction between electrons can be generated in the system.
See also Section~\ref{s:MeanField} where we treat the superconducting paring field.
\section{Proof}
\label{s:Proof}
\textit{Proof of Proposition~\ref{proposition}.}
In the following, 
we assume that the
conditions \eqref{eq:condition_X} and \eqref{eq:condition_mu} are
satisfied. 
We also assume that the electron number 
$\Ne$ is fixed.

Firstly we shall show that the Hamiltonian $H$ can be expressed as 
a sum of positive semi-definite operators. 
We define  $a$ operators by 
\begin{equation}
 \as{x}=
    w_{x+1} \cs{x} - w_{x} \cs{x+1}
\end{equation}
for $x\in\barLa$ and
\begin{equation}
 \as{L}=w_2\cs{1}+w_{L-1}\cs{L}.
\end{equation}
We also define $b$ operators by
\begin{equation}
 \bs{x}=
    w_{x} \cs{x} + w_{x+1} \cs{x+1}
\end{equation}
for $x\in\barLa$ and
\begin{equation}
 \bs{L}=-w_1\cs{1}+w_{L}\cs{L}.
\end{equation}
By using the $a$ operators and the $b$ operators we define
\begin{equation}
 H_{0,x} = V\left(\aupd{x}\bup{x}+\adnd{x}\bdn{x}\right)
           \left(\bupd{x}\aup{x}+\bdnd{x}\adn{x}\right).
\end{equation} 
It is noted that $H_{x,0}$ is positive semi-definite.
Then, after a lengthy but straightforward calculation, one finds that
$H_x$ is rewritten as
\begin{equation}
  H_x=H_{0,x}+H_{U^\prime,x}+H_{J^\prime,x}+H_{W,x},
   \label{eq:rewritten_H_x}
\end{equation}
where 
$H_{U^\prime,x}$ and $H_{J^\prime,x}$  
are, respectively, 
defined by \eqref{eq:on-site} and \eqref{eq:spin-spin}
with $U$ and $J$ replaced by $U^\prime=U-U_0$ and $J^\prime=J-J_0$, 
and 
$H_{W,x}$ is defined by 
\begin{eqnarray}
 H_{W,x}   & = & W(\cd{x,\up}\cd{x,\dn}+\cd{x+1,\up}\cd{x+1,\dn})
  (c_{x,\dn}c_{x,\up}+c_{x+1,\dn}c_{x+1,\up})
\end{eqnarray}
with $W=U_0/2$.
For $U\ge U_0$ and $J\ge J_0$, all the terms 
in the right hand side of \eqref{eq:rewritten_H_x} are positive semi-definite.
This proves that $H$ is the sum of the positive semi-definite operators
for $U>U_0$, 
$J\ge J_0$.
Therefore, 
a zero energy state of $H$, if it exists, is a ground state.

Let us next show that $\PhiG$ is a zero energy state 
of all the terms in \eqref{eq:rewritten_H_x} for any $x\in\barLa$.

Note that 
the $\tilde{a}$ operators form a basis for fermion operators
on $\La$, 
since $\{ \tasd{x},\as{y} \}=\delta_{x,y}$ for $x,y\in\La$ 
by our definition.
So we expand $\bs{x}$ with $x\in\La$ in terms of $\tas{x}$
as 
$
 \bs{x}=\sum_{y\in\La} \{\asd{y}, \bs{x}\}\tas{y}
$.
It is easy to see that $\{\asd{y}, \bs{x}\}={\sf F}_{y,x}$, which
gives us
\begin{equation}
 \bs{x}=\sum_{y\in\La} {\sf F}_{y,x}\tas{y} .
\end{equation}
From this expression of the $b$ operators we obtain
\begin{eqnarray}
 \aup{x}
\left(\sum_{y,z\in\La}{\sf F}_{y,z}\taupd{y} \taupd{z}\right)
&=&
 \left(
  \sum_{z\in\La} {\sf F}_{x,z} \taupd{z}
	  -\sum_{y,z\in\La}{\sf F}_{y,z}\taupd{y}\as{x} \taupd{z}
 \right)
\nonumber\\
&=&
 \left\{
  -\bupd{x}
	  -\sum_{y\in\La}{\sf F}_{y,x}\taupd{y}
  +\left(\sum_{y,z\in\La}{\sf F}_{y,z}\taupd{y} \taupd{z}\right)
    \aup{x}\right\}
\nonumber\\
&=&
 \left\{
  -2\bupd{x}
  +\left(\sum_{y,z\in\La}{\sf F}_{y,z}\taupd{y} \taupd{z}\right)
    \aup{x}\right\}.
\label{eq:a_zeta}
\end{eqnarray}
Since $(\bupd{x})^2=0$,
\eqref{eq:a_zeta} implies 
that
$\zeta_{\up,\up}^\dagger$ commutes with
$(\bupd{x}\aup{x}+\bdnd{x}\adn{x})$ for  $x\in\barLa$.
The creation operator
$\taupd{L}$ anticommutes with $\aup{x}$, 
i.e, it also commutes 
with $(\bupd{x}\aup{x}+\bdnd{x}\adn{x})$ for $x\in\barLa$.
Therefore, we have $H_{0,x}\PhiG=0$. 
This together with the fact
that there is no creation operator with the $\dn$-spin in $\PhiG$
leads to $H_x\PhiG=0$ for any $x\in\barLa$. 
This proves that $\PhiG$ is a zero energy state of $H$.     
From the $c$ operator representation of $\PhiG$ (see Appendix~\ref{s:groundstatec}), 
we find that the ground state is not the null state.

Finally we shall show the uniqueness of the zero energy state.

Let $M$ be the eigenvalue of the third component of the total spin.  
Since the Hamiltonian $H$ has the spin-rotation symmetry,
it is convenient to decompose the Hilbert space ${\mathcal H}$ of states 
into the subspaces ${\mathcal H}_M$ each of which has the fixed eigenvalue $M$.
Let $\Phi_{M}$ be a lowest-energy state in ${\mathcal H}_M$.
Since the representative of $\PhiG$ in ${\mathcal H}_M$ is also 
a zero energy state of $H$,
the lowest energy in ${\mathcal H}_M$ is guaranteed to be zero.
This implies that
$\Phi_M$ must satisfy $H_x\Phi_M=0$ for $x\in\barLa$.
In particular, for $U>U_0$, 
$c_{x,\dn}c_{x,\up}\Phi_M$ must be zero
for any $x\in\La$. 
Now we represent $\Phi_M$ by using the  $c$ operators. 
As mentioned above, since each site is forbidden to be doubly occupied by electrons in $\Phi_M$, 
it can be expanded in terms of normalized  basis states in the form
\begin{equation}
 \left(\prod_{x\in A} \cd{x,\sigma_x}\right)\Phi_0,
 \label{eq:basis}
\end{equation}
where $A$ is a subset of $\La$ with $|A|=\Ne$,
$\sigma_x=\up,\dn$, and $\sum_{x\in A}\sigma_x=M$.
In the product, the $c$ operators are ordered in such a way that
the site indexes $x$ increase from left to right.

Let us consider the matrix representation ${\sf H}$ of the Hamiltonian $H$ 
with respect to the basis states in the form \eqref{eq:basis}.
We  assume that the basis states are ordered in an arbitrary manner
and denote by ${\sf H}_{i,j}$ 
the matrix element corresponding to $i$-th and $j$-th basis states. 
Then one easily finds that 
any non-zero off-diagonal matrix element 
is $-t$ or $-J/2$, which is negative.
It is also easy to see that
for any $i,j$ there is a sequence $i_1,i_2,\dots,i_k$ such that
${\sf H}_{i,i_1}{\sf H}_{i_1,i_2}\dots{\sf H}_{i_k,j}\ne 0$. 
Therefore it follows from the Perron-Frobenius theorem that
the lowest energy state of ${\sf H}$ is unique~\cite{Tasaki98},
which implies
that the lowest energy state of $H$ in ${\mathcal H}_M$ 
is also unique and is given by
the representative of $\PhiG$ in ${\mathcal H}_M$.
This completes the proof of Proposition \ref{proposition}.

Before ending this section, we make a remark 
on the related exact results of the extended Hubbard model.
In the above proof we have rewritten the Hamiltonian as a sum of 
the positive semi-definite operators, 
and then have shown that the ground state attains the lowest eigenvalue, zero, 
of these operators.
This strategy was used in Ref.~\cite{SV95} to determine 
a parameter range for which the extended Hubbard model has the ferromagnetic 
ground states at half-filling (where the number of electrons is equal to that of sites).
Our result corresponds to an extension of Ref.~\cite{SV95} to the case away from half-filling.
Note, however, that our method for the construction of the exact ground state away from half-filling
is quite different from that at half-filling.
%
\section{Ground states with broken spin-rotation and gauge symmetries}
\label{s:GSwithBrokenSymmetries}
In this section, we assume that the parameters $U,J,X_x$ and $\mu_x$ 
satisfy the conditions in Proposition~\ref{proposition}, and hence
the ground states of $H$ with the fixed electron number 
are given by \eqref{eq:ground_states} and its SU(2) rotations.
Since the ground states are saturated ferromagnetic, 
we furthermore assume that 
the third component of the total spin is fixed
to $\Ne/2$. 
In the following,
since all the electrons are assumed to have the $\up$-spin,
we omit the spin indexes in the fermion operators for notational simplicity. 

The spin-triplet electron pairing ground state of $H$ is  
regarded as the pairing state of spinless fermions. 
The ground state of our model is thus expected to have 
some similar aspects to that of the Kitaev chain model in which
there appears the Majorana edge state at the ends of the chain. 
We will show that it is the case.

Let us define the zero energy ground states with the broken gauge symmetry 
\begin{equation}
 \PhiGe=
  \exp \left(
		-\frac{\eta}{2}\rme^{-\im\theta}\zeta^\dagger
	       \right)\Phi_0
 \label{eq:PhiGe}
\end{equation}
and
\begin{equation}
 \PhiGo=\sqrt{2\eta\sin\delta}\,\tadL\exp \left(
		  -\frac{\eta}{2}\rme^{-\im\theta}\zeta^\dagger
		   \right)\Phi_0
        = \sqrt{2\eta\sin\delta}\,\tadL\PhiGe,
 \label{eq:PhiGo}
\end{equation}
where $\eta$ is a positive parameter and $\theta$ is a phase parameter
(note that $\zeta^\dagger=\zeta_{\up,\up}^\dagger$).
The state $\PhiGe(\PhiGo)$ is a superposition of the zero energy states of $H$ 
with even(odd) numbers of electrons.  
The states $\PhiGe$ and $\PhiGo$ have the different fermionic parities,
and, as we shall see in the next section, these states are stabilized by superconducting
pairing fields.

As usual, let us define the Majorana fermion operators 
\begin{eqnarray}
  \gamma_{A,x}&=&\rme^{\im\frac{\theta}{2}}c_{x}
                 +\rme^{-\im\frac{\theta}{2}}\cd{x},\\
  \gamma_{B,x}&=&-\im\rme^{\im\frac{\theta}{2}}c_{x}
                   +\im\rme^{-\im\frac{\theta}{2}}\cd{x},
\end{eqnarray}
which satisfy $\gamma^\dagger_{\alpha,x}=\gamma_{\alpha,x}$
and
$\{\gamma_{\alpha,x},\gamma_{\beta,y}\}=2\delta_{\alpha,\beta}\delta_{x,y}$
for any $\alpha,\beta\in\{A,B\}$ and $x,y\in\La$. 
By using $\gamma_{\alpha,1}$ and $\gamma_{\alpha,L}$ with $\alpha=A,B$,
we introduce new edge operators as
\begin{eqnarray}
 \Gamma_1 & = & \frac{1}{\sqrt{2\eta\sin\delta}}
           \left\{
	    \left(w_2+\eta w_1 \right)\gamma_{A,1}
	    +\im\left(w_2-\eta w_1 \right)\gamma_{B,1}
           \right\},
 \\
 \Gamma_L & = & \frac{1}{\sqrt{2\eta\sin\delta}}
           \left\{
	    \left(w_2+\eta w_1 \right)\gamma_{B,L}
	    -\im\left(w_2-\eta w_1 \right)\gamma_{A,L}
           \right\}.
\end{eqnarray}
(Recall that $w_1=\sin(\delta/2)$ and $w_2=\cos(\delta/2)$.)
The edge operators $\Gamma_1$ and $\Gamma_L$ are rewritten as
\begin{eqnarray}
\Gamma_{1}&=&
 \sqrt{\frac{2}{\eta\sin\delta}} 
 \left(
  w_2 \rme^{\im\frac{\theta}{2}}
  c_1
  +
   \eta w_1\rme^{-\im\frac{\theta}{2}}
   c_1^\dagger
 \right)\\
 \Gamma_{L}&=&\im\sqrt{\frac{2}{\eta\sin\delta}} 
 \left(
  -w_2\rme^{\im\frac{\theta}{2}} c_L
  +
  \eta w_1\rme^{-\im\frac{\theta}{2}}
   c_L^\dagger
\right)
\end{eqnarray}
with the $c$ operators.
Then, we find that 
\begin{eqnarray}
 &&\Gamma_1 \Phi_{\mathrm{G},0} 
  = -\im\Gamma_L \Phi_{\mathrm{G},0}
  = \rme^{-\im\frac{\theta}{2}} \Phi_{\mathrm{G},1},
   \label{eq:Gamma1PhiG0}\\
 &&\Gamma_1 \Phi_{\mathrm{G},1} 
  = \im\Gamma_L \Phi_{\mathrm{G},1}
  = \rme^{\im\frac{\theta}{2}} \Phi_{\mathrm{G},0}.
  \label{eq:Gamma1PhiG1}
\end{eqnarray}
Furthermore, from the above relations, we obtain  
\begin{eqnarray}
 -\im\Gamma_1\Gamma_L \Phi_{\mathrm{G},0} &=& \Phi_{\mathrm{G},0}, \\
 -\im\Gamma_1\Gamma_L \Phi_{\mathrm{G},1} &=& -\Phi_{\mathrm{G},1}.
\end{eqnarray}

The relations \eqref{eq:Gamma1PhiG0} and \eqref{eq:Gamma1PhiG1} are
obtained as follows.
For $x\in\La$  we have from \eqref{eq:a_zeta} that
\begin{equation}
 \rme^{\im\frac{\theta}{2}}a_{x}\left(-\frac{\eta}{2}\rme^{-\im\theta}\zeta^\dagger\right)^n\Phi_0
  =\eta\rme^{-\im\frac{\theta}{2}}\bd{x}n\left(-\frac{\eta}{2}\rme^{-\im\theta}\zeta^\dagger\right)^{n-1}\Phi_0,
\end{equation}
which yields
\begin{equation}
 \rme^{\im\frac{\theta}{2}}a_{x}\PhiGe=
  \eta\rme^{-\im\frac{\theta}{2}}b^\dagger_{x}\PhiGe.
\label{eq:aPhiGe}
\end{equation}
Here we used 
$\bd{x}\left(\zeta^\dagger\right)^\frac{L-1}{2}\Phi_0=0$~\cite{comment1}.
By \eqref{eq:aPhiGe}, we also have
\begin{equation}
 \rme^{\im\frac{\theta}{2}}a_{x}\PhiGo=
  \rme^{\im\frac{\theta}{2}}a_{x} \left(\sqrt{2\eta\sin\delta}\,\tadL\PhiGe\right)
 =
  \eta\rme^{-\im\frac{\theta}{2}}b^\dagger_{x}\PhiGo
 +\delta_{x,L}\sqrt{2\eta\sin\delta}\,\rme^{\im\frac{\theta}{2}}\PhiGe.
  \label{eq:aPhiGo}
\end{equation}
By representing \eqref{eq:aPhiGe} and \eqref{eq:aPhiGo} with the $c$
operators and setting  $x=L$, one finds
\begin{eqnarray}
   \left\{
    w_2\rme^{\im\frac{\theta}{2}} 
    \left(
     c_1
     +
     c_L
     \right)
     +  
     \eta w_1\rme^{-\im\frac{\theta}{2}}
     \left(
      c_1^\dagger
      -  
     c_L^\dagger
    \right)
   \right\}
   \Phi_{\mathrm{G},l}
   = \delta_{l,1} \sqrt{2\eta\sin\delta}\,\rme^{\im\frac{\theta}{2}}\Phi_{\mathrm{G},0}
   \label{eq:edge1}
\end{eqnarray}
with $l=0,1$.
On the other hand, \eqref{eq:aPhiGe} and \eqref{eq:aPhiGo}
combined with 
\begin{eqnarray}
   \sum_{x\in\barLa} \bd{x} & = &-w_1\cd{1}-w_1\cd{L}+2\sin\delta \tadL \\
   \sum_{x\in\barLa} a_x   & = &  w_2 c_{1}- w_2 c_{L},
\end{eqnarray}
which follow from the definition, yield
\begin{eqnarray}
\left\{
 w_2\rme^{\im\frac{\theta}{2}} 
 \left(
  c_{1}-c_{L}
  \right)
  +\eta w_1\rme^{-\im\frac{\theta}{2}}
  \left(
   \cd{1}+\cd{L}
  \right)
\right\}
\Phi_{\mathrm{G},l}
 =\delta_{l,0}\sqrt{2\eta\sin\delta}\rme^{-\im\frac{\theta}{2}}\PhiGo
\label{eq:edge2}
\end{eqnarray}
with $l=0,1$.
From \eqref{eq:edge1} and \eqref{eq:edge2} we obtain \eqref{eq:Gamma1PhiG0},
and \eqref{eq:Gamma1PhiG1}.
 
It is noted that in the case $\eta=w_2/w_1=1/\tan(\delta/2)$ we have
$\Gamma_1=\gamma_{A,1}$ and $\Gamma_L=\gamma_{B,L}$ which are the
Majorana fermion operators.
In this case, we can reconstruct the edge fermion operator by combining
$\gamma_{A,1}$ and $\gamma_{B,L}$ as
\begin{equation}
 d_\mathrm{edge} = \frac{1}{2} 
\rme^{-\im\frac{\theta}{2}}\left(\gamma_{A,1}
+\im\gamma_{B,L}\right).
\end{equation}
The fermion operator $d_\mathrm{edge}$ satisfies 
$\{d_\mathrm{edge},d_\mathrm{edge}\}
=\{d_\mathrm{edge}^\dagger,d_\mathrm{edge}^\dagger\}=0$ 
and 
$\{d_\mathrm{edge}^\dagger,d_\mathrm{edge}\}=1$.
From
\eqref{eq:Gamma1PhiG0} and \eqref{eq:Gamma1PhiG1} we also have
\begin{eqnarray}
 d_\mathrm{edge}^\dagger \PhiGe = \PhiGo, \\
 d_\mathrm{edge} \PhiGo = \PhiGe. 
\end{eqnarray}
The above relations yield $n_\mathrm{edge}\PhiGo=\PhiGo$ and
$n_\mathrm{edge}\PhiGe=0$ with $n_\mathrm{edge}=d_\mathrm{edge}^\dagger d_\mathrm{edge}$, 
which imply that the Majorana edge state is formed
at the ends of the chain.
\section{Mean field Hamiltonian}
\label{s:MeanField}
In this section we consider external fields (or mean fields) which
remove the ground state degeneracy and select $\PhiGe$ and $\PhiGo$
as the two ground states.

It is well known that the external magnetic field  
can remove the degeneracy due to the spin-rotation symmetry.
So we assume that the system is in a magnetic field,
and fix the third component of the total spin to $\Ne/2$.
(As in the previous section, the spin indexes are omitted in this and the next sections
under this assumption.
)

In order to remove the degeneracy due to the electron pair condensation, we shall consider
the Hamiltonian which does not conserve the electron number.
More precisely, we will introduce Hamiltonian $H^\prime$ of spinless fermions 
with superconducting pairing field, 
and show that the ground states of $H+H^\prime$ are given by $\PhiGe$ and $\PhiGo$.

Let us define
\begin{eqnarray}
H^\prime
 & = & \sum_{x\in\barLa} H_x^\prime, \\
H_x^\prime 
 &=& 
   \frac{|\Delta|}{\eta}
    (\rme^{-\im\frac{\theta}{2}}a^\dagger_{x}
  -\eta\rme^{\im\frac{\theta}{2}}b_{x})
  (\alpha + (1-\alpha)a_{x} a_{x}^\dagger)
  (\rme^{\im\frac{\theta}{2}}a_{x}
  -\eta\rme^{-\im\frac{\theta}{2}}b^\dagger_{x}
  ),
\end{eqnarray} 
where $\alpha$ and $|\Delta|$ are non-negative parameters.
As we will see below, $\Delta=|\Delta|\rme^{\im\theta}$ corresponds 
to the superconducting pairing field.
Since $\{a_x^\dagger,a_x\}=1$ for $x\in\barLa$, we have
\begin{equation}
 H_x^\prime 
 = 
   \frac{|\Delta|}{\eta}
    (\rme^{-\im\frac{\theta}{2}}a^\dagger_{x}
  -\eta\rme^{\im\frac{\theta}{2}}b_{x})
  (1 - (1-\alpha)a_{x}^\dagger a_{x} )
  (\rme^{\im\frac{\theta}{2}}a_{x}
  -\eta\rme^{-\im\frac{\theta}{2}}b^\dagger_{x}
  ),
\end{equation}
and hence $H_x^\prime$ is a positive semi-definite
operator for $\alpha\ge 0$.
From \eqref{eq:aPhiGe} and \eqref{eq:aPhiGo} we find that
$\PhiGe$ and $\PhiGo$ are zero energy states of $H_x^\prime$ for
$x\in\barLa$.
Therefore $\PhiGe$ and $\PhiGo$ are ground states of $H+H^\prime$.
It is easy to see that there is no other ground state.
The Hamiltonian $H^\prime$ removes the ground state degeneracy
of $H$ and stabilizes the states $\PhiGe$ and $\PhiGo$.

After some lengthy but straightforward calculations, $H^\prime$ is rewritten as
\begin{eqnarray}
 H^\prime 
 &=&
 -s\sum_{x\in\barLa}(c_{x}^\dagger c_{x+1}+c_{x+1}^\dagger c_x)
 -\sum_{x\in\barLa}(\nu_x c_x^\dagger c_x +\nu_{x+1}c_{x+1}^\dagger c_{x+1})
 \nonumber\\
 &&
  -V^\prime \sum_{x\in\barLa} c_x^\dagger c_x c_{x+1}^\dagger c_{x+1}
  +\sum_{x\in\barLa} (\Delta c_{x}c_{x+1}
  +\Delta^\ast c_{x+1}^\dagger c_{x}^\dagger )
  +\eta|\Delta|(L-1)
\end{eqnarray}
with
\begin{eqnarray}
 s & = & \frac{|\Delta|}{2\eta}
                  (1+\alpha\eta^2)
               \sin \delta, \\
 \nu_x & = & -\frac{|\Delta|}{2\eta}
  \left\{(1+\alpha\eta^2-2\eta^2)
         -(-1)^x(1+\alpha\eta^2)\cos \delta\right\}, \\
 V^{\prime} & = &  (\alpha-1)\eta|\Delta|. 
\end{eqnarray}
From the above representation of $H^\prime$, one immediately realizes
that $\Delta$ corresponds to the superconducting pairing field,
which may be induced from a nearby superconductor.
This field term essentially removes the degeneracy.
It is noted that, 
in the case where $\delta=\pi/2$, $\eta=1$ and $\alpha=1$,
$H^\prime$ is reduced to the Hamiltonian of 
the Kitaev chain of the spinless fermions in the topological phase.
Thus our model can be also regarded as an extension of the Kitaev chain
to the spinful system with the electron-electron interactions. 
%
\section{Electron number conserving case}
\label{s:ElectronNoConserving}
In the previous two sections we considered 
the case where the number of electrons
is not conserved. 
From the expressions \eqref{eq:PhiGe} and \eqref{eq:PhiGo}
of the symmetry breaking ground states,
one finds that the edge state is closely related to the zero energy mode
corresponding to $\tadL$.
Indeed, we have shown that the occupation of $\tadL$ by an electron is reflected as an eigenvalue
of the number operator $n_\mathrm{edge}$ of the edge fermion operator.

For the fixed electron number, the ground state $\PhiG$ can not be the eigenstate of $n_\mathrm{edge}$,
since we have 
\begin{equation}
 n_\mathrm{edge}=\frac{1}{2}\left(1+ \im \gamma_{A,1}\gamma_{B,L}\right)
\end{equation}
with
\begin{equation}
 \im \gamma_{A,1}\gamma_{B,L} = \rme^{\im\theta}c_1 c_L + \rme^{-\im\theta} c_L^\dagger c_1^\dagger
                                       +c_1^\dagger c_L + c_L^\dagger c_1.
\end{equation}
Instead, we can expect that there is a difference between the expectation values 
of $n_\mathrm{edge}$ for $\PhiG$ with $\Ne$ even and odd. 

Let $\expece{\cdots}$ and $\expeco{\cdots}$ be the expectation values 
$\langle \PhiG, \cdots \PhiG\rangle/\langle \PhiG,\PhiG\rangle$
for $\PhiG$ with $\Ne$ even and odd, respectively.
We will estimate $\expece{n_\mathrm{edge}}$ and $\expeco{n_\mathrm{edge}}$.
Clearly, we have $\expecl{c_1 c_L}=\expecl{c_L^\dagger c_1^\dagger}=0$ 
and $\expecl{c_1^\dagger c_L}= \expecl{c_L^\dagger c_1}$ with $l=0,1$.
Let us consider $\expecl{c_1^\dagger c_L}$.
By using the $c$ operator representation of $\PhiG$ (see Appendix~\ref{s:groundstatec}), we obtain
\begin{eqnarray}
 \expecl{c_1^\dagger c_L} 
 = 
 (-1)^{l+1}\sin^2\left(\frac{\delta}{2}\right)
 \frac{\sum_{A\subset \La;|A|=\Ne-1} \chi[1,L\notin A]~W_A}
  {\sum_{A\subset \La;|A|=\Ne}W_A}.
\end{eqnarray}
where $W_A=\prod_{x\in A}w_x^2$, and $\chi[E]$ takes the value 1 if $E$ is true and 0 otherwise.  
Since we have 
\begin{eqnarray}
 \sum_{A\subset \La;|A|=\Ne}W_A \le 
\frac{L(L-1)}{\Ne(L-\Ne)} \cos^2\left(\frac{\delta}{2}\right)\sum_{A\subset \La;|A|=\Ne-1} \chi[1,L\notin A] W_A
\label{eq:inequality}
\end{eqnarray}
(see Appendix~\ref{s:inequality}),
$|\expecl{c_1^\dagger c_L}|$ is bounded from below as
\begin{eqnarray}
 |\expecl{c_1^\dagger c_L}|\ge \tan^2\left(\frac{\delta}{2}\right)\rho(1-\rho)
\end{eqnarray}
with $\rho=\Ne/L$.
Therefore, we obtain
\begin{eqnarray}
 \expece{n_\mathrm{edge}} &\le& \frac{1}{2}-\tan^2\left(\frac{\delta}{2}\right)\rho(1-\rho) 
\label{eq:nedge0}
\\
 \expeco{n_\mathrm{edge}} &\ge& \frac{1}{2}+\tan^2\left(\frac{\delta}{2}\right)\rho(1-\rho). 
\label{eq:nedge1}
\end{eqnarray}

The inequalities obtained above show that 
the ground state expectation value of the occupation number $n_\mathrm{edge}$
corresponding to the edge fermion reconstructed 
by the Majorana fermions depends 
on the fermionic parity, regardless of the chain length $L$. 
Let $\Ne$ be even.
Since we have $\rho=\Ne/L\approx(\Ne+1)/L\approx(\Ne-1)/L$ for sufficiently large $L$,
these inequalities imply that the expectation value of $n_\mathrm{edge}$ for $\Ne$ decreases
by at least $2\tan^2(\delta/2)\rho(1-\rho)$ compared with that for $\Ne-1$, 
while the expectation value of $n_\mathrm{edge}$ for $\Ne+1$ increases by at least $2\tan^2(\delta/2)\rho(1-\rho)$
compared with that for $\Ne$.  
This behavior in the edge fermion number 
indicates the formation of an edge state
in the electron number conserving setting. 
In the following, 
we propose a concrete example of a system having the two-fold degenerate ground states
each of which is characterized by a zero energy mode related to the Majorana edge state.
Very recently, a similar model has been investigated in Refs.~\cite{LB15} and \cite{IMRFD15}.

Firstly we prepare a copy of $H$. 
The operators in the copied system are denoted 
by the underline as $\underline{c}_{x}$.   
We then consider the Hamiltonian $H+\underline{H}+H_\epsilon$ on the two chains,
where
\begin{equation}
 H_{\epsilon} = \epsilon\left\{\sum_{\sigma=\up,\dn}(\underline{a}_{1,\sigma}^\dagger b_{1,\sigma} 
                                    + a_{1,\sigma}^\dagger \underline{b}_{1,\sigma})\right\}
                        \left\{\sum_{\sigma=\up,\dn}(b_{1,\sigma}^\dagger \underline{a}_{1,\sigma}
                                    + \underline{b}_{1,\sigma}^\dagger a_{1,\sigma} )\right\}
\end{equation} 
with $\epsilon>0$ is an interchain interaction. 
The number of electrons on the whole system 
is fixed to $\Ne$.
We suppose that the values of the parameters in $H$ and $\underline{H}$
are taken so that each Hamiltonian is positive semi-definite 
and has the zero energy ground states (see Proposition~\ref{proposition}).  
Under the assumption that the system is in a magnetic field,
one finds that 
the two states 
\begin{eqnarray}
\PhiGe^\prime 
=
\tadL\left(\zeta^\dagger+\underline{\zeta}^\dagger\right)^\Np \Phi_0,~
\PhiGo^\prime
=
\utadL\left(\zeta^\dagger+\underline{\zeta}^\dagger\right)^\Np \Phi_0
\end{eqnarray}
for odd $\Ne$, and
\begin{eqnarray}
\PhiGe^\prime 
=
\left(\zeta^\dagger+\underline{\zeta}^\dagger\right)^\Np \Phi_0,~
\PhiGo^\prime
=
\tadL\utadL\left(\zeta^\dagger+\underline{\zeta}^\dagger\right)^{\Np-1} \Phi_0
\end{eqnarray} 
for even $\Ne$ 
are the only ground states of this system.
In fact, $H$, $\underline{H}$ and $H_\epsilon$ are positive semi-definite, 
and $\PhiGe^\prime$ and $\PhiGo^\prime$ are the only zero energy states 
for these Hamiltonians.
It is expected that similar inequalities corresponding to \eqref{eq:nedge0} 
and \eqref{eq:nedge1} hold for 
the ground state expectation values of 
the number operators $n_\mathrm{edge}$ and $\underline{n}_\mathrm{edge}$ of 
the edge fermions on the chains,
although explicit analytical expressions are difficult to obtain.

We end this section with the remark
 that the fermion operator defined by $a_{\pi}=\sum_{x\in\La}(-1)^{x+1}a_{x}$
plays an interesting role in manipulating the zero energy mode in the condensate.
More precisely, $a_{\pi}$ satisfies $a_\pi\zeta^\dagger = \zeta^\dagger a_{\pi}$
since $\sum_{x\in\La}(-1)^{x+1}b^\dagger=0$ and $\{\tadL,a_\pi\}=\{\tadL,a_L\}=1$. 
Therefore, we have the relations
$(\sqrt{2\eta\sin\delta})^{-1}a_\pi\PhiGo=\PhiGe$, $\tadL a_\pi \PhiGo=\PhiGo$ and 
$\tadL a_\pi \PhiGe=0$
for the symmetry breaking ground states. 
Similar relations are also found for the electron number 
conserving system. Namely, we have 
$\tadL a_{\pi} \PhiGe^\prime = \PhiGe^\prime$,
$\utadL \underline{a}_{\pi} \PhiGo^\prime = \PhiGo^\prime$ and 
$\utadL \underline{a}_{\pi} \PhiGe^\prime =\tadL {a}_{\pi} \PhiGo^\prime = 0$ for odd $\Ne$,
and  
$\tadL {a}_{\pi}\PhiGo^\prime=\utadL\underline{a}_{\pi} \PhiGo^\prime=\PhiGo^\prime$
and 
$\tadL {a}_{\pi} \PhiGe^\prime=\utadL\underline{a}_{\pi} \PhiGe^\prime=0$
for even $\Ne$.
\section{Spin-Spin Interaction with Ising-like Anisotropy}
\label{s:Anisotropy}
In this section, we treat the case of the spin-spin interaction with
an Ising-like anisotropy.

Let us define
\begin{equation}
     \left(\bm{S}_x\cdot \bm{S}_{x+1}\right)_\beta
      =S_{x}^{(3)}S_{x+1}^{(3)}
      +\beta(S_{x}^{(1)}S_{x+1}^{(1)}+S_{x}^{(2)}S_{x+1}^{(2)})
\end{equation}
where $\beta$ is a non-negative parameter
and denote by $H_{J,\beta,x}$ 
the Hamiltonian obtained by replacing $\bm{S}_x\cdot \bm{S}_{x+1}$
with $\left(\bm{S}_x\cdot \bm{S}_{x+1}\right)_\beta$ in $H_{J,x}$.
Then we consider the Hamiltonian
\begin{eqnarray}
 H_{\beta} & = & \sum_{x\in\barLa} H_{\beta,x}, \\
 H_{\beta,x} & = & H_{t,x}+H_{U,x} +H_{V,x}+H_{J,\beta,x}.
\end{eqnarray} 
Note that the bond-charge interaction $H_{X,x}$ is omitted
in $H_{\beta,x}$. 
For $H_\beta$, we have the following result:
\begin{proposition}
\label{proposition2}
 Suppose that both $U>U_0+2X_0$ and $J> J_0+4X_0$ are satisfied.
 We furthermore suppose that $\mu_x$ is given by \eqref{eq:condition_mu}.
 Then, the ground state energy of $H_\beta$ with 
 \begin{equation}
   \frac{J_0}{J}\le \beta < 1-\frac{4X_0}{J}
 \end{equation}
 is zero.
 For fixed $\Ne$, the ground state 
 is two-fold degenerate and is given by
 \begin{equation}
  \PhiG=
   \left\{
   \begin{array}{@{\,}ll}
     \left(\zeta^\dagger_{\sigma,\sigma}\right)^\Np\Phi_0 
      & \mbox{for even $\Ne$} \\
     \tasd{L}\left(\zeta^\dagger_{\sigma,\sigma}\right)^\Np\Phi_0 
      & \mbox{for odd $\Ne$} \\
   \end{array}
   \right.
\label{eq:ground_states_with_anisotropy}
 \end{equation}
with $\sigma=\up,\dn$.
\end{proposition}

The outline of the proof is as follows.
As in the isotropic spin-spin interaction case, we rewrite $H_{\beta,x}$
as
\begin{equation}
 H_{\beta,x}=H_{0,x}+H_{U^{\prime\prime},x}+H_{J^{\prime\prime},\beta^\prime,x}
                    +H_{X_0,x}+H_{W,x},
\label{eq:anisotropic_hamiltonian}
\end{equation}
where $H_{X_0,x}$ is given by
\begin{eqnarray}
 H_{X_0,x}
  &=&  X_0\sum_{\sigma=\up,\dn}
  \left\{c_{x,\sigma}^\dagger+(-1)^{x}c_{x+1,\sigma}^\dagger\right\}
  n_{x,-\sigma}\left\{c_{x,\sigma}+(-1)^{x}c_{x+1,\sigma}\right\}
  \nonumber\\
 &&  + X_0\sum_{\sigma=\up,\dn} 
  \left\{c_{x,\sigma}^\dagger-(-1)^{x}c_{x+1,\sigma}^\dagger\right\}
  n_{x+1,-\sigma}\left\{c_{x,\sigma}-(-1)^{x}c_{x+1,\sigma}\right\},
\end{eqnarray}
$H_{U^{\prime\prime},x}$ is obtained by replacing $U$ with
$U^{\prime\prime}=U-U_0-2X_0$ in $H_{U,x}$, and 
$H_{J^{\prime\prime},\beta^\prime,x}$ is obtained by replacing $J$ and $\beta$ with
$J^{\prime\prime}=J-J_0-4X_0$ and $\beta^\prime=(J\beta-J_0)/(J-J_0-4X_0)$,
respectively,
in $H_{J,\beta,x}$.  
When $U^{\prime\prime}>0,J^{\prime\prime}>0$ and $0\le\beta^\prime<1$,
all the terms in $\eqref{eq:anisotropic_hamiltonian}$
are positive semi-definite and $\PhiG$ 
in \eqref{eq:ground_states_with_anisotropy} is their zero energy state.
The fact that there is no other zero energy state follows from
the application of the Perron-Frobenius theorem. 

In the case of the isotropic spin-spin interaction, 
the bond-charge interaction whose strength parameter is fixed
must be included in the Hamiltonian to obtain the exact ground states.
On the other hand, 
the Hamiltonian with the anisotropic spin-spin interaction
has the exact ground states even if   
the bond-charge interaction is absent.
Although the on-site potentials still have to be adjusted
to  certain values, 
the model with the anisotropic spin-spin interaction exhibits 
the spin-triplet electron pair condensation 
over the wide range of parameters.
\section{Conclusion}
\label{s:Conclusion}
We have introduced the one-dimensional extended Hubbard model whose
ground state simultaneously 
exhibits saturated ferromagnetism and spin-triplet electron pair
condensation under certain conditions.
Recently, the extended Hubbard chain with
charge-charge and spin-spin interactions at low filling 
was studied 
by means of mean field and numerical methods
in Ref.~\cite{SCHW14}.
The results showed
that 
the ground state is in the spin-triplet pairing phase 
for strong ferromagnetic coupling, 
even if there are no fine-tuned bond-charge interactions
and on-site potentials which are necessary to get our exact results.  
These results together with ours indicate that 
the model exhibits 
spin-triplet pairing over a wide range of parameters.
We have constructed two ground states in which 
both of the spin-rotation symmetry and the gauge symmetry are broken.
It has been shown that these ground states are transferred from one to the other
by applying the edge operators. 
The edge operators become the Majorana fermions in a certain case,
and, in this sense, the Majorana state is formed on the edges of a chain 
in our model. 
We have introduced the mean field Hamiltonian with the pairing field 
which stabilizes 
the gauge symmetry breaking ground states.
Here we remark that the spin-triplet pair condensation found in the ground state of $H$
is unstable against the thermal fluctuation since $H$ is constituted of short-range interactions
and is defined on a chain.
However we can expect that the spin-triplet pair condensate survives at non-zero temperatures 
in the strong pairing field.
The mean field Hamiltonian has been
shown to be regarded as the Kitaev chain with 
the nearest neighbour charge-charge interaction.
It is noted that a similar spinless fermion model 
has been studied by Katsura, Schuricht, and Takahashi recently~\cite{KST15}.
Our extended Hubbard model together with the mean field
is an extension of the Kitaev chain to the spinful electron model.
We have also estimated the expectation values of the edge fermion number operator 
for the ground states with fixed even and odd numbers of electrons,
and found that there is the difference between them.
Furthermore, we have proposed the model on the two chains 
in the electron number conserving setting and 
have shown that the model has the two-fold degenerate ground states
which are characterized by the zero modes on the chains.

To conclude, it is interesting to note 
that Nadj-Perge \textit{et al.} reported 
the observation of Majorana fermions in a chain of Fe atoms,
which intrinsically have ferromagnetic nature,
on a superconducting Pb substrate~\cite{Lee14,Nadj14}. 
It is also noted that the recent developments 
in the field of cold atoms open
a route to the experimental realization of 
one-dimensional interacting fermion systems~\cite{Guan13}.
We hope that our results stimulate these fields.

\section*{Acknowledgements}
I would like to thank H. Katsura and M. Takahashi 
for valuable discussion.
I also would like to thank K. Sun for bringing his work to my attention.
This work was supported by JSPS KAKENHI Grant Numbers 
25400407 and 25287076. 

\begin{appendices}
\renewcommand{\theequation}{\thesection.\arabic{equation}}
\section{$c$ operator representation of $\PhiG$}
\label{s:groundstatec}
\setcounter{equation}{0}
Here we express $\PhiG$ in terms of the $c$ operators. 
For the notational simplicity, we omit the spin indexes of the fermion operators.  

From \eqref{eq:Fxy} and \eqref{eq:zeta}, one obtains
\begin{eqnarray}
 \zeta^\dagger  =  -\sin\delta\left(\sum_{x\in\barLa}\tad{x} \tad{x+1}+\tad{1}\tad{L}\right).
\label{eq:zetaa}
\end{eqnarray}
Substituting \eqref{eq:ta} into the right hand side of \eqref{eq:zetaa}, we have 
\begin{eqnarray}
\zeta^\dagger=  -\frac{4}{\sin\delta}\sum_{x,y\in\La, x<y} (w_x w_{y} c_{x}^\dagger c_{y}^\dagger).
\end{eqnarray}
Then,
taking into account the sign factor arising from the exchange of fermion operators,
we have
\begin{equation}
 (\zeta^\dagger)^\Np\Phi_0
=
\left(-\frac{4}{\sin\delta}\right)^\Np(\Np!)\left(
 \sum_{A\subset \La;|A|=2\Np}\prod_{x\in A}(w_x c_x^\dagger)\right)\Phi_0
\end{equation}
and
\begin{equation}
 \tad{L}(\zeta^\dagger)^\Np\Phi_0
=
\frac{1}{\sin\delta}\left(-\frac{4}{\sin\delta}\right)^\Np(\Np!)
\left(\sum_{A\subset \La;|A|=2\Np+1}
\prod_{x\in A}(w_x c_x^\dagger)\right)\Phi_0
\end{equation}
where $\Np!=\Np(\Np-1)\cdots2\cdot1$, and $|A|$ denotes the number of elements in a set $A$.
\section{Proof of the inequality \eqref{eq:inequality}}
\label{s:inequality}
\setcounter{equation}{0}
Let us prove the inequality \eqref{eq:inequality}.
Firstly we rewrite the left hand side as
\begin{eqnarray}
 \sum_{A\subset\La;|A|=\Ne} W_A= \frac{1}{\Ne}\sum_{x\in\La}w_x^2\sum_{A\subset\La;|A|=\Ne-1} \chi[x\notin A]~W_A.
\end{eqnarray}
Then, by using $w_L=\sin(\delta/2)\le w_x \le \cos(\delta/2)$ (recall that $0<\delta\le\pi/2$), 
we obtain
\begin{eqnarray}
 \sum_{A\subset\La;|A|=\Ne} W_A &\le& \frac{1}{\Ne}\cos^2\left(\frac{\delta}{2}\right)\sum_{x\in\La}\sum_{A\subset\La;|A|=\Ne-1} \chi[x\notin A]~W_A 
\nonumber\\
&\le& \frac{L}{\Ne}\cos^2\left(\frac{\delta}{2}\right)\sum_{A\subset\La;|A|=\Ne-1} \chi[L\notin A]~W_A. 
\end{eqnarray}
Here, note that $\chi[L\notin A]=\chi[1,L\notin A]+\chi[1\in A,L\notin A]$.
Since the sum related to $\chi[1\in A,L\notin A]$ is bounded as 
\begin{eqnarray}
 \lefteqn{\sum_{A\subset\La;|A|=\Ne-1} \chi[1\in A,L\notin A]~W_A} 
\nonumber\\
&=& \sin^2\left(\frac{\delta}{2}\right)\sum_{A\subset\La;|A|=\Ne-2} \chi[1,L\notin A]~W_A 
\nonumber\\
&=& \sin^2\left(\frac{\delta}{2}\right)\sum_{A\subset\La;|A|=\Ne-2} \chi[1,L\notin A]~W_A\sum_{x\in\La} \frac{w_x^2}{w_x^2(L-\Ne)}\chi[x\notin A\cup\{1,L\}] 
\nonumber\\
&\le&
\frac{1}{L-\Ne}
\sum_{A\subset\La;|A|=\Ne-2}\sum_{x\in\La}  \chi[1,L\notin A]\chi[x\notin A\cup\{1,L\}]~W_A w_x^2
\nonumber\\
&=&
\frac{\Ne-1}{L-\Ne}
\sum_{A\subset\La;|A|=\Ne-1}  \chi[1,L\notin A]W_A,
\end{eqnarray}
we conclude that
\begin{eqnarray}
 \sum_{A\subset\La;|A|=\Ne} W_A 
  &\le& 
  \frac{L}{\Ne}\left(1+\frac{\Ne-1}{L-\Ne}\right)\cos^2\left(\frac{\delta}{2}\right)
  \sum_{A\subset\La;|A|=\Ne-1}  \chi[1,L\notin A]~W_A
\nonumber\\
  &=&
  \frac{L}{\Ne}\left(\frac{L-1}{L-\Ne}\right)\cos^2\left(\frac{\delta}{2}\right)
  \sum_{A\subset\La;|A|=\Ne-1}  \chi[1,L\notin A]~W_A.
\end{eqnarray}
\end{appendices}


\begin{thebibliography}{99}
%
 \bibitem{Hirsch93}
 J. E. Hirsch,
 Phys. Rev. Lett. \textbf{53} (1984) 2327.
%
 \bibitem{JK99}
 G. I. Japaridze and A. P. Kampf,
 Phys. Rev. B \textbf{59} (1999) 12822.
%
 \bibitem{TML07}
N. Toyota, J. M\"{u}ller and M. Lang,
\textit{Low-Dimensional Molecular Metals}
(Springer-Verlag, Berlin Heidelberg, 2007).
%
 \bibitem{EKS92}
 F. H. L. Essler, V. E. Korepin and K. Schoutens,
 Phys. Rev. Lett. \textbf{68} (1992) 2960.
%
 \bibitem{AA94}
 L. Arrachea and A. A. Aligia,
 Phys. Rev. Lett. \textbf{73} (1994) 2240.
%
 \bibitem{DM01}
 D. Dolcini and A. Montorsi,
 Nucl. Phys. B \textbf{592} (2001) 563.
%
 \bibitem{Tanaka08}
 A. Tanaka,
 J. Phys. A: Math. Theor. \textbf{41} (2008) 365208.
%
 \bibitem{Kitaev01}
 A. Yu Kitaev,
 Phys. Usp. \textbf{44} (2001) 131.
%
 \bibitem{Fendley12}
 P. Fendley,
 J. Stat. Mech. (2012) 11020
%
 \bibitem{LSS10}
 R. M. Lutchyn, J. D. Sau, and S. Das Sarma,
 Phys. Rev. Lett. \textbf{105} (2010) 077001
%
 \bibitem{Alicea11}
 J. Alicea, Y. Oreg, G. Refael, F. von Oppen and M. P. A. Fisher,
 Nature Physics \textbf{7} (2011) 412.
%
 \bibitem{Mourik12}
 V. Mourik, K. Zuo, S. M. Frolov, S. R. Plissard, 
 E. P. A. M. Bakkers and  L. P. Kouwenhoven,
 Science \textbf{336} (2012) 1003.
%
 \bibitem{Lee14}
 P. A. Lee,
 Science \textbf{346} (2014) 545.
%
 \bibitem{Nadj14}
 S. Nadj-Perge, I. K. Drozdov, J. Li, H. Chen, S. Jeon, J. Seo,
 A. H. MacDonald, B. A. Bernevig and A. Yazdani,
 Science \textbf{346} (2014) 602.
%
 \bibitem{Tasaki98}
 H. Tasaki,
 Prog. Theor. Phys. \textbf{99} (1998) 489.
%
 \bibitem{SV95}
 R. Strack and D. Vollhardt,
 J. Low Temp. Phys. \textbf{99} (1995) 385.
%
 \bibitem{comment1}
 Since $\bd{x}\left(\zeta^\dagger\right)^\frac{L-1}{2}\Phi_0$ is
a state with $L$ electrons, we have 
$\bd{x}\left(\zeta^\dagger\right)^\frac{L-1}{2}\Phi_0
=C \prod_{y\in\La} \tilde{a}_{y}^\dagger\Phi_0$ 
with a certain constant number $C$.
Note that $\{\bd{x},a_{x}\}=0$.
Then, by applying $a_{x}$ on the both sides of this equation,
we find that $C=0$.
%
 \bibitem{LB15}
 N. Lang and H. P. B\"uchler,
 Phys. Rev. B \textbf{92} (2015) 041118(R).
%
 \bibitem{IMRFD15}
F. Iemini, L. Mazza, D. Rossini, R. Fazio and S. Diehl,
Phys. Rev. Lett. \textbf{115} (2015) 156402.
%
 \bibitem{SCHW14}
 K. Sun, C.-K. Chiu, H.-H. Hung and  J. Wu,
 Phys. Rev. B \textbf{89} (2014) 104519.
%
 \bibitem{KST15}
 H. Katsura, D. Schuricht and  M. Takahashi,
 Phys. Rev. B \textbf{92} (2015) 115137.
%
 \bibitem{Guan13}
 X.-W. Guan, M. T. Batchelor and C. Lee,
 Rev. Mod. Phys. \textbf{85} (2013) 1633.
\end{thebibliography}
\end{document}